\documentclass[conference]{IEEEtran}
\usepackage{cite}
\usepackage{amsmath,amssymb,amsfonts}
\usepackage[normalem]{ulem}
\usepackage{enumitem}
\usepackage{graphicx}
\usepackage{booktabs}
\usepackage{mathtools}
\usepackage{subcaption}
\usepackage{xcolor}
\usepackage{soul}

\IEEEoverridecommandlockouts
\begin{document}
\author{
\IEEEauthorblockN{Alireza Mohammadhosseini and Fatemeh Afghah}\thanks{This material is based upon work supported by the National Science Foundation under Grant Numbers CNS-2202972, CNS-2318726, and CNS-2232048}
\IEEEauthorblockA{\textit{Department of Electrical and Computer Engineering, Clemson University, Clemson, SC, USA}}}

\title{Place, Slice and Schedule: Hierarchical O-RAN Control of a Tethered mmWave UAV-gNB}

\maketitle

\begin{abstract}
Unmanned aerial vehicle (UAV)-mounted 5G New Radio base stations (gNBs) can augment terrestrial networks with an on-demand, repositionable Frequency Range 2 (FR2) capacity layer. This flexibility, however, couples the physical network topology with radio-resource management: UAV movement reshapes blockage, channel quality, and the set of effectively served users, while traffic demand, queues, and service requirements evolve at a much faster timescale. Existing Open Radio Access Network (O-RAN)-enabled UAV studies optimize trajectory, deployment, association, or resource allocation, but typically in isolation, without coordinating slow aerial control with fast per-user scheduling. We instead exploit O-RAN disaggregation, Key Performance Indicator (KPI) monitoring, and multi-timescale RAN Intelligent Controller (RIC) control to address this coupling: a Non-Real-Time RIC rApp uses aggregated KPIs and radio-environment context to jointly control tethered UAV placement and the enhanced Mobile Broadband (eMBB)/Ultra-Reliable Low-Latency Communication (URLLC) slice budget, while a Near-Real-Time RIC xApp allocates per-user resources within that budget. We realize this xApp as a permutation-equivariant \emph{DeepSets Soft Actor-Critic (D-SAC) scheduler} that treats the users as an unordered set, trained in a Sionna RT ray-traced channel. The resulting hierarchical controller improves eMBB SLA satisfaction by up to $17\%$ and URLLC on-time delivery by up to $42\%$ over classical and learned schedulers; the learned rApp further raises URLLC on-time delivery by up to $20\%$ over baselines.
\end{abstract}

\section{Introduction}
\label{sec:intro}

UAV-mounted base stations have emerged as a flexible way to add capacity and coverage on demand~\cite{ref_uav_survey}, and tethered UAVs have been introduced to provide persistent power and remove the flight-time limit~\cite{tuav_coverage}. Serving such aerial cells well requires programmable, adaptive control of both where the UAV is placed and how it allocates radio resources to meet the service requirements of 5G-advanced and 6G networks. Open Radio Access Network (O-RAN) provides this through a disaggregated, programmable architecture whose control logic is exposed by the RAN Intelligent Controller (RIC)~\cite{ref_oran}: control is organized by timescale, with a non-real time RIC (Non-RT RIC) hosting rApps that generate high-level policy over the A1 interface at second-level timescales, and a near-real time RIC (Near-RT RIC) hosting xApps that act over E2 within $10$~ms--$1$~s. O-RAN is emerging as a programmable control architecture for non-terrestrial RAN, enabling RIC-based optimization of mobility, radio resources, and service continuity across multiple timescales~\cite{oran_ntn}.

These decisions span timescales: slow placement and slice-budget choices shape coverage and long-term service satisfaction, while fast per-user allocation decides whether individual service-level agreements (SLAs) are met. Because a decision at one layer propagates to the other and changes quality-of-service (QoS) or reliability outcomes~\cite{telecom_wm}, the two loops cannot be tuned independently, especially at millimeter-wave frequencies, where the UAV position determines per-user coverage through line-of-sight blockage.
\begin{figure}[t]
    \centering
    \includegraphics[width=\linewidth]{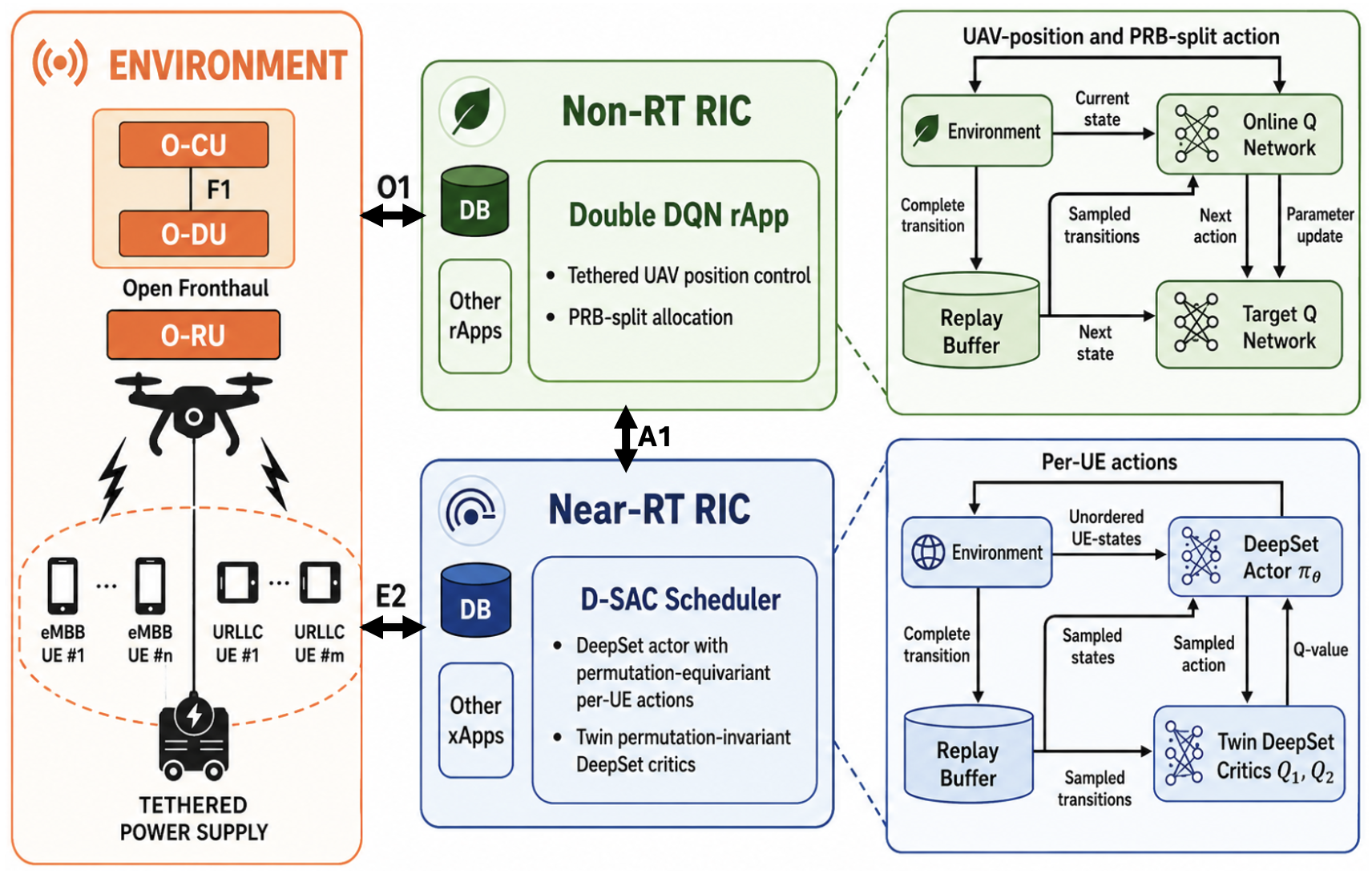}
    \caption{Hierarchical O-RAN control of the tethered mmWave UAV-gNB: the rApp sets UAV 3D-placement and the slice PRB budget, and the D-SAC xApp allocates PRBs per user.}
\label{fig:sys}
\end{figure}

A second challenge arises from the structure of near-real-time resource control. As the number of active UEs grows, a flat multilayer perceptron (MLP) policy is tied to a fixed input dimension and UE ordering, although UE indices have no physical meaning and the active-user population may vary over time; such policies neither preserve scheduling decisions under UE permutations nor transfer to different user counts without retraining. Graph-based RL~\cite{xslice} can handle variable-size observations, but requires constructing an explicit user graph and message passing, even when users interact only through competition for a common resource budget. We therefore model the active UE population as an unordered set and design a permutation-equivariant, parameter-shared policy, which provides structural scalability while retaining a fixed architecture as the number and ordering of UEs change.

We propose a hierarchical O-RAN controller for a tethered UAV-gNB that co-designs both loops, illustrated in Fig.~\ref{fig:sys}. The Non-RT rApp selects the tethered UAV placement together with the eMBB/URLLC slice PRB split, handed to the Near-RT xApp as an A1 policy; conditioned on that policy, the xApp performs per-user PRB allocation every radio slot. We realize the xApp as a permutation-equivariant \emph{DeepSets SAC scheduler} (D-SAC)~\cite{ref_deepsets,ref_SAC} and decouple the loops with a staged procedure: the scheduler is trained under randomized high-level policy, the rApp is trained over the frozen scheduler, and the scheduler is fine-tuned to the rApp policy. Channel gains for every tethered UAV candidate are precomputed with Sionna RT ray tracing~\cite{ref_sionna}, capturing the line-of-sight and blockage geometry that drives mmWave coverage rather than relying on a stochastic channel model. Our contributions are:
\begin{itemize}[leftmargin=*]
\item A hierarchical two-timescale intelligent O-RAN controller for a tethered mmWave UAV-gNB in which a rApp sets UAV 3D-placement and slice PRB budget and a xApp performs conditional per-user scheduling, together with a staged co-design procedure that decouples the two RIC loops.
\item A scalable permutation-equivariant DeepSets Soft Actor-Critic (D-SAC) scheduler for variable-cardinality UE populations. Shared per-UE processing and permutation-invariant global pooling remove dependence on UE indexing, enable a single policy architecture to operate across different user counts, and avoid the explicit graph construction required by graph-based schedulers.
\item A geometry-aware evaluation using Sionna RT ray-traced FR2 channels demonstrating that the hierarchical controller improves both eMBB SLA satisfaction and URLLC on-time delivery over classical and learned baselines, while the same D-SAC policy retains its performance advantage as the UE population increases without retraining.
\end{itemize}

\section{Related Work}
\label{sec:related}
UAV placement has been studied for guaranteed mmWave line-of-sight coverage~\cite{mmwave_los}, for integrated access and backhaul~\cite{drl_placement_iab}, and for joint positioning and association via deep Q-learning~\cite{drl_ra_uav}, while tethered UAVs have been analyzed for coverage~\cite{tuav_coverage} and controlled via multi-agent Q-learning~\cite{tuav_maql}. However, UAV placement is commonly optimized with respect to coverage, association, throughput, or another aggregate network objective while the fast per-user scheduler is fixed or abstracted. This misses an important coupling: each UAV movement changes the channel state and feasible service region available to the scheduler, while heterogeneous user traffic and service-level requirements determine which UAV positions are desirable in the first place.

The O-RAN architecture provides a programmable framework for exposing these coupled control variables through RIC loops operating at different timescales. U-ORAN~\cite{uoran} introduced RIC-based control for multi-UAV networks and jointly optimized UAV trajectories and task offloading, while~\cite{oran_uav_ee} jointly considered aerial radio-unit deployment, user association, and resource allocation in an O-RAN-enabled UAV network. These works demonstrate the value of O-RAN for intelligent aerial-network control, but the radio-resource decisions remain primarily coupled to trajectory, association, or offloading objectives rather than to a service-aware per-user scheduler operating at a distinct faster control timescale. Separately, terrestrial O-RAN research has investigated learning-based radio control: xSlice~\cite{xslice} employs actor--critic deep RL with a graph representation to support a varying number of traffic sessions, CollabORAN~\cite{collaboran} coordinates resource control across multiple RIC timescales, and~\cite{perue_ran} considers learning-based per-UE RAN parameter adaptation, while full-stack xApp implementations on srsRAN demonstrate closed-loop RL slicing~\cite{real_xapp} and meta-RL improves adaptation to nonstationary O-RAN traffic~\cite{meta_hrl}. These studies establish the feasibility of multi-timescale and intelligent control in O-RAN, but do not address the additional feedback loop introduced when the gNB itself is a controllable aerial platform.

When UEs mainly compete for a shared resource budget, a set-based representation provides a simpler alternative to graph-based scheduling: we represent the active UE population as an unordered set and use a permutation-equivariant DeepSets policy whose encoder and decision layers are shared across users and whose global context is obtained through permutation-invariant pooling, yielding a single parameterization that is independent of UE ordering and operates across different UE counts. Different from the above lines of work, we jointly address the two structural challenges introduced by an aerial O-RAN: \emph{multi-timescale coupling between physical topology and radio-resource control}, and \emph{variable-cardinality per-user scheduling}.

\section{System Model}
\label{sec:system_model}

We consider a tethered UAV-gNB serving mobile eMBB and URLLC users over a ray-traced mmWave downlink, controlled by two coupled O-RAN layers: a slow Non-RT rApp that selects the UAV 3D-placement and the slice-level PRB budget, and a fast Near-RT xApp that allocates the slice PRB budgets among individual UEs conditioned on that policy. In our O-RAN system, the rApp generates the A1 policy, and the xApp enforces its per-UE PRB decisions through E2SM-RC while consuming per-slot KPIs via E2SM-KPM over E2. The rApp state is built from O1 performance-management telemetry, supplied as R1 enrichment information.

\subsection{Channel Model}
\label{subsec:channel}
The UAV-gNB is tethered to a fixed ground anchor and can only occupy positions in a feasible tethered candidate set $\mathcal{V}$. A generic tethered position can be parameterized by the cable length $L_{\mathrm{te}}$, an elevation angle $\rho$ measured from the vertical axis, and an azimuth angle $\phi$:
\begin{equation}
\label{eq:tether_position}
  \mathbf{v}(\rho,\phi)
  =
  \mathbf{v}_{0}
  + L_{\mathrm{te}}
  \begin{bmatrix}
  \sin\rho\cos\phi \\
  \sin\rho\sin\phi \\
  \cos\rho
  \end{bmatrix},
\end{equation}
where $\mathbf{v}_{0}$ is the ground anchor. In the control problem, the rApp selects a neighboring candidate $\mathbf{v}_{t+1}\in\mathcal{N}_{\mathrm{te}}(\mathbf{v}_t)$ from the tether-feasible navigation graph.

Let $\mathbf{x}_{u,t}$ be the position of UE $u$ at step $t$, $\mathbf{v}_{t}\in\mathcal{V}$ the selected UAV candidate, and $\beta(\mathbf{v}_{t},\mathbf{x}_{u,t})$ the location-dependent power gain obtained from the Sionna RT radio map. The system has $N$ PRBs of bandwidth $B_{\mathrm{PRB}}$. We model a single UAV-gNB downlink cell, so link quality is represented by the downlink SNR of each UE. With uniform power allocation across PRBs, the transmit power per PRB is $P_{\mathrm{tx}}/N$, and the downlink SNR of UE $u$ in slot $\ell$ of rApp step $t$ is:
\begin{equation}
  \label{eq:snr}
  \gamma_{u,t,\ell}
  =
  \frac{(P_{\mathrm{tx}}/N)
        \beta(\mathbf{v}_{t},\mathbf{x}_{u,t})
        h_{u,t,\ell}}
       {N_{0}B_{\mathrm{PRB}}},
\end{equation}
where $P_{\mathrm{tx}}$ is total transmit power, $h_{u,t,\ell}$ is an independent unit-mean small-scale fading power gain capturing slot-scale variations not represented by the precomputed radio map, and $N_{0}$ is the noise spectral density including receiver noise figure. If UE $u$ receives $n_{u,t,\ell}$ PRBs, its downlink rate is $R_{u,t,\ell}= n_{u,t,\ell} B_{\mathrm{PRB}}\log_{2}(1+\gamma_{u,t,\ell})$.

\subsection{Service Model and QoS Metrics}
\label{subsec:service}

The UAV-gNB serves eMBB UEs $\mathcal{U}_{e}$ and URLLC UEs $\mathcal{U}_{u}$, with $\mathcal{U}_{e}\cup\mathcal{U}_{u}=\mathcal{U}$. The rApp selected budget $N_t^e$ limits the total eMBB PRBs available in control slot~$t$, while the xApp decides the per-UE allocations $n_{u,t,\ell}$ at each radio slot. The mean eMBB throughput over the $L$ radio intervals of control window~$t$ is $\bar{R}_{t}^{e}=\frac{1}{L|\mathcal{U}_{e}|}\sum_{\ell=0}^{L-1}\sum_{u\in\mathcal{U}_{e}}R_{u,t,\ell}$, which we map to a bounded eMBB QoS score:
\begin{equation}\label{eq:embb-qos}
  q_t^e=\min\left(\frac{\bar{R}_t^e}{R_{\mathrm{target}}},1\right),
\end{equation}
where $R_{\mathrm{target}}$ is a network-level reference rate that normalizes this aggregate reward-shaping term. The eMBB SLA is enforced per UE, with a violation whenever a UE's throughput falls below its own minimum-rate guarantee $R_{u,\min}$. The eMBB violation ratio is:
\begin{equation}
  \label{eq:embb_viol}
  \mathcal{V}_{t}^{e}
  = \frac{1}{L|\mathcal{U}_{e}|}
    \sum_{\ell=0}^{L-1}\sum_{u\in\mathcal{U}_{e}}
    \mathbb{I}\!\left[R_{u,t,\ell}<R_{u,\min}\right].
\end{equation}

URLLC is modeled as packet-based, deadline-constrained GBR traffic with per-UE packet delay budgets in the tens of milliseconds, and on-time delivery is reported as a scheduling metric under an intentionally overloaded regime rather than as a reliability figure. For URLLC UE~$u$, packet arrivals in radio interval $\ell$ follow $A_{u,t,\ell}\sim\operatorname{Poisson}(\lambda_{u}\Delta)$, where $\lambda_{u}$ is the arrival rate and $\Delta$ is the radio slot duration. Packets are queued per UE and may be retransmitted while their deadlines and retransmission budgets permit, so the queued bits evolve as:
\begin{equation}
  \label{eq:queue}
  Q_{u,t,\ell+1}
  = \max (0, Q_{u,t,\ell}+b_{u}A_{u,t,\ell}-S_{u,t,\ell}),
\end{equation}
where $b_u$ is packet size in bits and $S_{u,t,\ell}$ is the service in bits, bounded by capacity $C_{u,t,\ell}=R_{u,t,\ell}\,\Delta$ and by the packet decoding outcome. A packet misses its deadline if it is not delivered within $D_{u}$ seconds of arrival, giving the window-level deadline-miss rate $\mathcal{V}_{t}^{u}=N_{t}^{\mathrm{miss}}/\max(1,N_{t}^{\mathrm{del}}+N_{t}^{\mathrm{miss}})$, where $N_{t}^{\mathrm{del}}$ and $N_{t}^{\mathrm{miss}}$ are the number of delivered and missed packets in window~$t$. With the mean URLLC queue $\bar{Q}_{t}^{u}=\frac{1}{L|\mathcal{U}_{u}|}\sum_{\ell}\sum_{u\in\mathcal{U}_{u}}Q_{u,t,\ell}$, we define URLLC QoS as:
\begin{equation}
  \label{eq:queue_qos}
  q_{t}^{u}=1-\frac{\bar{Q}_{t}^{u}}{\bar{Q}_{t}^{u}+Q_{\mathrm{ref}}},
\end{equation}
where $Q_{\mathrm{ref}}$ is the queue level at which the queue-pressure term reaches one half. Thus $q_{t}^{u}$ provides QoS feedback before deadlines expire, while $\mathcal{V}_{t}^{u}$ measures SLA violation.

\section{Problem Formulation}
\label{subsec:opt}

We formulate the controller as a hierarchical MDP, splitting control into a slow rApp decision and a fast xApp scheduling decision. The rApp acts only on O-RAN aggregate KPIs and does not require per-slot per-UE MAC (Media Access Control) state, whereas the xApp acts on the finer per-UE radio-control information.

\subsection{Hierarchical MDP}
\label{subsec:hmdp}

At rApp step $t$, the slow-timescale MDP is $\mathcal{M}_{R}=(\mathcal{S}_{R},\mathcal{A}_{R},P_R,r_R,\Gamma_R)$. The rApp observes $s_t^R\in\mathcal{S}_{R}$ and chooses a joint action $a_t^R=(\alpha_t,\eta_t)\in\mathcal{A}_R$, where $\alpha_t$ is a tethered UAV movement command and $\eta_t=(\eta_t^e,\eta_t^u)\in\mathcal{H}$ is the eMBB/URLLC slice PRB split from a finite split set $\mathcal{H}$. The movement command applies a single-axis step $\alpha_t\in\{\mathtt{stay},\,\rho^{+},\,\rho^{-},\,\phi^{+},\,\phi^{-}\}$; these primitives generate exactly the tether-feasible neighborhood $\mathcal{N}_{\mathrm{te}}(\mathbf{v}_t)$ of~\eqref{eq:tether_position}, restricting movement to one grid hop per control step. The chosen split fixes the per-slice PRB budgets $N_t^e=\eta_t^eN$ and $N_t^u=\eta_t^uN$ for the following rApp window. The state $s_t^R$ is a fixed-length vector of window-aggregated features: the normalized UAV placement and previous movement, the previous slice split, per-slice UE spatial and radio-quality summaries, QoS-pressure and slice-composition statistics, and a candidate-lookahead block summarizing the per-slice service each neighboring position would provide. All entries are aggregate measurements over the previous $1\,\mathrm{s}$ window, consistent with the KPI-level information a Non-RT RIC observes rather than per-slot MAC scheduler state.

Within the rApp window, the xApp problem is modeled as the MDP $\mathcal{M}_{X}(a_t^R)=(\mathcal{S}_{X},\mathcal{A}_{X},P_X,r_X,\Gamma_X\mid a_t^R)$. At radio slot $\ell$, it observes a per-UE feature matrix $\mathbf{F}_{t,\ell}=[{f}_{1,t,\ell},\ldots,{f}_{|\mathcal{U}|,t,\ell}]^{\mathsf{T}}$, where each row $f_{u,t,\ell}\in\mathbb{R}^{12}$ combines the channel, queue, throughput, and deadline/SLA state of UE $u$ with the rApp slice-budget context. The xApp emits one continuous action per UE, $\mathbf{a}_{t,\ell}^{X}=[a_{1,t,\ell}^{X},\ldots,a_{|\mathcal{U}|,t,\ell}^{X}]^{\mathsf{T}}$, which is mapped to a feasible PRB allocation $\mathbf{n}_{t,\ell}$ inside the rApp selected slice budgets. For each slice $s\in\{e,u\}$:
\vspace{-0.5cm}
\begin{equation}
\label{eq:slice_feasible}
  \sum_{u\in\mathcal{U}_{s}}n_{u,t,\ell}\leq N_t^s,\qquad n_{u,t,\ell}\geq0.
\end{equation}
This continuous PRB-share relaxation represents average PRB shares over the radio-control interval. Integer PRB rounding is outside the present model. The resulting hierarchical policy factorizes as:

\begin{equation}
\label{eq:policy_factorization}
  \pi(a_t^R,\mathbf{a}_{t,0:L-1}^X)
  =
  \pi_R(a_t^R|s_t^R)
  \prod_{\ell=0}^{L-1}
  \pi_X(\mathbf{a}_{t,\ell}^X|\mathbf{F}_{t,\ell},a_t^R),
\end{equation}
which makes the dependency explicit: the xApp policy is conditioned on the rApp intent through the slice-budget context and the resulting channel/queue state. In both tuples, $P$ is the induced transition kernel and $\Gamma\in(0,1)$ the discount factor.

\subsection{Objective and Reward}

The window-level QoS score is the weighted sum of eMBB QoS in~\eqref{eq:embb-qos} and URLLC QoS in~\eqref{eq:queue_qos}, $U_t=w_e^{\mathrm{qos}}q_t^e+w_u^{\mathrm{qos}}q_t^u$. There is no movement-energy term because the UAV is tethered and powered through the cable; mobility is constrained through the feasible tether graph instead of penalized by an energy model. For an episode horizon of $T$ rApp steps, the joint hierarchical objective is:
\begin{subequations}
\label{eq:opt}
\begin{align}
\max_{\pi_R,\pi_X}\quad
&\frac{1}{T}\sum_{t=0}^{T-1}
\mathbb{E}_{\pi_R,\pi_X}[U_t]
\label{eq:obj}\\
\text{s.t.}\quad
&\frac{1}{T}\sum_{t}
\mathbb{E}[\mathcal{V}_t^e]\leq\epsilon_e,\quad
\frac{1}{T}\sum_{t}
\mathbb{E}[\mathcal{V}_t^u]\leq\epsilon_u,\\
&\mathbf{v}_{t+1}\in\mathcal{N}_{\mathrm{te}}(\mathbf{v}_t),\quad
\eta_t^e+\eta_t^u=1,\\
&\eta_t^e,\;\eta_t^u,\;n_{u,t,\ell}\geq 0,\quad \forall u,t,\ell,
\end{align}
\end{subequations}
where $\mathcal{N}_{\mathrm{te}}(\mathbf{v}_t)$ is the tether-feasible neighbor set and $\epsilon_e,\epsilon_u$ denote SLA violation limits. For training, we relax the constrained problem into an unconstrained per-step reward with fixed penalty weights $\kappa_e,\kappa_u>0$, i.e. a Lagrangian relaxation with fixed multipliers:
\begin{equation}
  \label{eq:training_reward}
  r_t
  =
  U_t
  -\kappa_e\mathcal{V}_t^e
  -\kappa_u\mathcal{V}_t^u.
\end{equation}

\section{Proposed Solution}
\label{subsec:training_decomposition}

End-to-end optimization of \eqref{eq:opt} is computationally expensive because one rApp action induces many xApp decisions, so we use a staged training procedure that preserves the hierarchy in \eqref{eq:policy_factorization}. First, the xApp is trained as a conditional fast-timescale scheduler under randomized UAV candidates and randomized slice budgets. Second, the xApp is frozen and the rApp is trained on the induced slow-timescale MDP. Third, the xApp is fine-tuned with the learned rApp frozen, which adapts the near-RT scheduler to the state distribution induced by the rApp; all reported D-SAC results use this fine-tuned scheduler.

\subsection{Per-UE PRB Scheduling}

For fixed rApp state and action, the UAV candidate and slice budgets are fixed during the rApp control window, and the xApp controls only the per-UE allocation vector $\mathbf{n}_{t,\ell}=[n_{1,t,\ell},\ldots,n_{|\mathcal{U}|,t,\ell}]^{\mathsf{T}}$ for each slot $\ell$, subject to the per-slice budgets of~\eqref{eq:slice_feasible}. The conditional xApp training problem is:
\begin{equation}
\label{eq:xapp_subproblem}
\max_{\pi_X}\;\;
\mathbb{E}_{\pi_X}\left[\sum_{\ell=0}^{L-1}r_X(\mathbf{F}_{t,\ell},\mathbf{n}_{t,\ell})\mid s_t^R,a_t^R\right],
\end{equation}
where the learning reward $r_X$ is the fast timescale counterpart of \eqref{eq:training_reward}: it rewards eMBB QoS and URLLC queue QoS, while penalizing eMBB SLA violation and URLLC deadline miss to provide dense scheduling feedback.

\subsection{DeepSets SAC Scheduler (D-SAC)}

The xApp is trained with Soft Actor-Critic (SAC)~\cite{ref_SAC}. A flat MLP actor would tie parameters to UE indices, which is undesirable for scheduling because if the UE order is permuted, the output scores should be permuted in the same way. Therefore, we keep SAC as the learning algorithm but parameterize its actor and critic with \emph{DeepSets} networks~\cite{ref_deepsets}. Unlike a GNN-based scheduler~\cite{gnn}, this design does not require constructing an explicit graph or performing message passing, which suits our setting because the xApp must map an unordered UE set to per-UE PRB priority scores, requiring permutation equivariance rather than explicit relational inference.

The actor applies shared per-UE layers $\mathbf{h}_{u,t,\ell}=\phi_\theta({f}_{u,t,\ell})$ to each feature vector, then forms a permutation-invariant global context using mean and max pooling:
\begin{equation}
  \mathbf{c}_{t,\ell}=\psi_\theta\Big(\frac{1}{|\mathcal{U}|}\sum_{j\in\mathcal{U}}\mathbf{h}_{j,t,\ell},\;\max_{j\in\mathcal{U}}\mathbf{h}_{j,t,\ell}\Big).
\end{equation}
Mean pooling captures the average traffic and channel condition, while max pooling captures extreme cases such as highly urgent or backlogged UEs. A shared stochastic SAC head $g_\theta$ then maps each UE's embedding and the global context to a per-UE action $a_{u,t,\ell}=\tanh(\mu_{u,t,\ell}+\sigma_{u,t,\ell}\,\epsilon_{u,t,\ell})\in(-1,1)$, with $(\mu_{u,t,\ell},\sigma_{u,t,\ell})=g_\theta(\mathbf{h}_{u,t,\ell},\mathbf{c}_{t,\ell})$ and SAC reparameterization noise $\epsilon_{u,t,\ell}\sim\mathcal{N}(0,1)$. Because all per-UE layers are shared and the context is permutation invariant, the actor is \emph{permutation equivariant} for any UE permutation matrix. The bounded actions are mapped to strictly positive scheduling priorities by a centered softplus:
\begin{equation}
\label{eq:xapp_score}
  z_{u,t,\ell}=\operatorname{softplus}\!\left(a_{u,t,\ell}-\frac{1}{|\mathcal{U}|}\sum_{j\in\mathcal{U}}a_{j,t,\ell}\right),
\end{equation}
which keeps the priorities positive and differentiable while centering them so that the per-slice normalization below responds to relative rather than absolute action values. The priorities are normalized within each slice to produce feasible PRB shares:
\begin{equation}
  n_{u,t,\ell}
  =
  \begin{cases}
  \displaystyle \eta_t^eN\frac{z_{u,t,\ell}}{\sum_{j\in\mathcal{U}_e}z_{j,t,\ell}}, & u\in\mathcal{U}_e,\\[8pt]
  \displaystyle \eta_t^uN\frac{z_{u,t,\ell}}{\sum_{j\in\mathcal{U}_u}z_{j,t,\ell}}, & u\in\mathcal{U}_u.
  \end{cases}
\end{equation}
The critic evaluates the whole allocation with one scalar Q-value, so it must be \emph{permutation invariant}. Each SAC twin critic applies shared layers $\mathbf{y}_{u,t,\ell}=\varphi_\omega({f}_{u,t,\ell},a_{u,t,\ell})$ to each per-UE feature/action pair, then pools over UEs and predicts $Q_\omega(\mathbf{F}_{t,\ell},\mathbf{a}_{t,\ell})=q_\omega\big(\frac{1}{|\mathcal{U}|}\sum_{j\in\mathcal{U}}\mathbf{y}_{j,t,\ell},\max_{j\in\mathcal{U}}\mathbf{y}_{j,t,\ell}\big)$. Thus, the actor maps an unordered UE set to per-UE scores equivariantly, while the critic maps the unordered set of UE feature/action pairs to one invariant value.

\subsection{rApp Training over the Frozen xApp}

In the second stage, $\pi_X$ is frozen and absorbed into the environment: the frozen per-UE scheduler, together with UE mobility, packet arrivals, and channel variation, forms the transition kernel of the slow-timescale MDP $\mathcal{M}_R$ of Sec.~\ref{subsec:hmdp}. This lets the rApp be trained as a standard MDP over its aggregate O-RAN state, with Double DQN~\cite{ref_ddqn} over the discrete joint action set of tether-movement commands and slice-budget splits detailed in Sec.~\ref{sec:experiments}.

\section{Experimental Setup}
\label{sec:experiments}

The deployment area is $200\,\mathrm{m}\times200\,\mathrm{m}$ and the UAV-gNB is tethered to the map center by a fixed $100\,\mathrm{m}$ cable. The candidate set uses five elevation levels $\{0,12.5,25,37.5,50\}^{\circ}$ and five azimuth levels $\{0,72,144,216,288\}^{\circ}$, giving $21$ geometrically distinct positions over an altitude range of $64.3$--$100\,\mathrm{m}$; the corresponding channel-gain maps are precomputed with Sionna RT at $28\,\mathrm{GHz}$ (n257). At each step the rApp selects a slice split from $\{(0.9,0.1),(0.7,0.3),(0.5,0.5),(0.3,0.7),(0.1,0.9)\}$, and UEs move at $1\,\mathrm{m/s}$ under a uniform random walk. Remaining parameters are in Table~\ref{tab:params}.

eMBB UEs are assigned different base loads and minimum-rate SLAs to emulate the heterogeneous experienced data rate requirements of 5G eMBB service tiers, and URLLC UEs different arrival rates and deadlines, creating unequal per-UE service pressure within each slice. These wide ranges place the system in a demanding regime where some eMBB minimum-rate targets exceed the per-UE capacity of the cell and part of the service area falls in ray-traced blockage where the channel gain drops to the noise floor. Therefore, eMBB SLA satisfaction and URLLC on-time delivery are bounded by the SLA and coverage geometry; the residual headroom is what the two control loops compete over, and we report them as comparative metrics under identical conditions. All results report the mean and $95\%$ confidence interval over five random seeds. A per-slot allocation by the D-SAC scheduler takes $0.12\,$ms for $30$ users on a single CPU thread (about $1\%$ of the $10\,$ms radio-slot budget), confirming Near-RT feasibility.

\begin{table}[t]
\vspace*{5pt}
\centering
\caption{Simulation and training parameters.}
\label{tab:params}
\small
\begin{tabular}{@{}ll@{}}
\toprule
Parameter & Value \\
\midrule
Carrier frequency / band
    & $28$~GHz (n257) \\
PRBs / PRB bandwidth
    & $132$ / $1.44$~MHz \\
Tx power / rApp, xApp slots
    & $36$~dBm / $1$~s, $10$~ms \\
UEs / eMBB:URLLC ratio
    & $30,40,50$ / $0.6:0.4$ \\
Episode length
    & $100$ steps \\
eMBB $R_{\mathrm{target}}$ / $R_{u,\min}$
    & $10$ / $[2,70]$~Mbps \\
URLLC packet / arrival
    & $256$~B / $[50,800]$~pkt/s \\
URLLC deadline $D_u$
    & $[20,50]$~ms \\
Penalties $(\kappa_e,\kappa_u)$ / $\Gamma$
    & $(3.0,3.0)$ / $0.99$ \\
D-SAC encoder / context / critic
    & $64$ / $64$ / $256$ \\
DDQN layers / xApp,rApp LR
    & $[256,256]$ / $3{\times}10^{-4},10^{-4}$ \\
Training steps (Phase 1/2/3)
    & $300$k / $150$k / $100$k \\
Evaluation seeds / episodes
    & $5$ / $50$ \\
\bottomrule
\end{tabular}
\end{table}

\begin{figure}[t]
  \centering
  \begin{subfigure}[t]{0.48\linewidth}
    \centering
    \includegraphics[width=\linewidth]{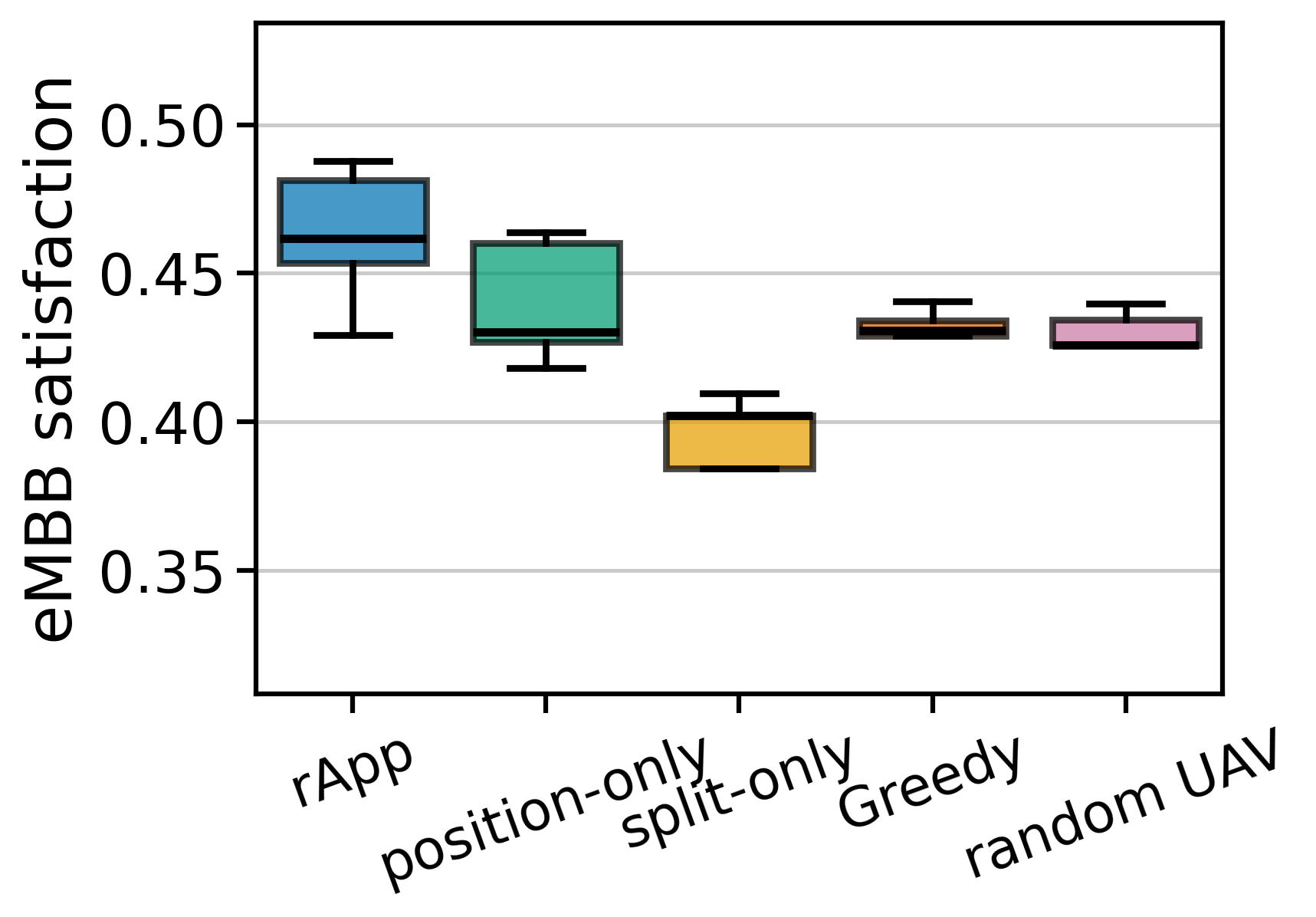}
    \caption{eMBB SLA satisfaction}
    \label{fig:results_rapp_embb}
  \end{subfigure}
  \hfill
  \begin{subfigure}[t]{0.48\linewidth}
    \centering
    \includegraphics[width=\linewidth]{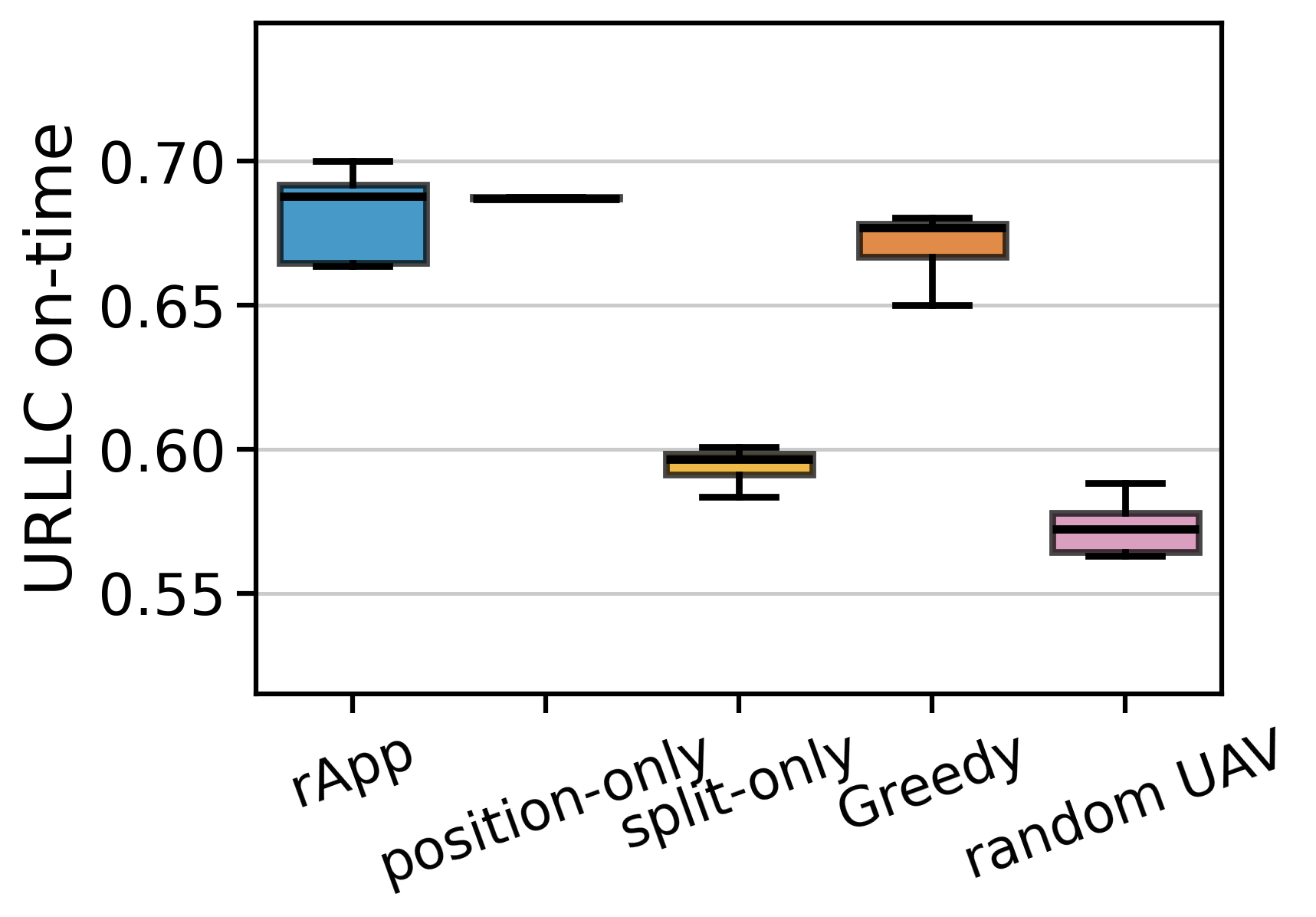}
    \caption{URLLC on-time delivery}
    \label{fig:results_rapp_urllc}
  \end{subfigure}
  \caption{Contribution of the rApp. The rApp improves eMBB SLA satisfaction and URLLC on-time delivery over baselines.}
  \label{fig:results_rapp}
\end{figure}

\section{Results}
\label{sec:results}
We evaluate the fully trained hierarchical controller of Sec.~\ref{subsec:training_decomposition}. To characterize service quality we report two per-UE metrics instead of raw violation rates. The eMBB SLA satisfaction ratio $
  \bar{s}^{e}=\frac{1}{|\mathcal{U}_e|}\sum_{u\in\mathcal{U}_e}\min\!\Big(\frac{\bar{R}_u}{R_{u,\min}},1\Big),
$ where $\bar{R}_u$ is UE $u$'s mean rate over the evaluation window, is the average fraction of each eMBB UE's contracted minimum rate that is delivered, and the URLLC on-time delivery ratio $\bar{o}^{u}=1-\mathcal{V}^{u}$ is the fraction of URLLC packets delivered within their deadline. We additionally report aggregate throughput and Jain's fairness~\cite{jain}.

\subsubsection{Performance Evaluation of the proposed rApp}
\label{subsec:results_rapp}
Fig.~\ref{fig:results_rapp} isolates the rApp by varying only the high-level policy while keeping the same low-level D-SAC scheduler. Alongside the full learned rApp we include two ablations that expose the marginal value of each rApp decision: \emph{rApp position-only} applies only the learned UAV placement decision with fixed slice split, and \emph{rApp split-only} applies only the learned slice split while holding the UAV at the map
center. We further compare against a \emph{greedy} method that at each step moves the UAV toward the reachable tether candidate with the best predicted cover
age from the radio map, and a random-movement UAV; both use a fixed split. Two effects stand out. First, learned placement is the dominant contributor: rApp move-only alone reaches $0.440$ eMBB satisfaction and $0.687$ URLLC on-time, far above the pinned rApp split-only ($0.391$, $0.594$), and the learned rApp surpasses even the strong greedy method on both axes. Second, learned slice control adds a distinct gain on top of placement: the full rApp lifts eMBB satisfaction from position-only $0.440$ to $0.463$ by shifting budget toward eMBB, trading only a negligible URLLC on-time ($0.682$ versus $0.687$) for that improvement. Relative to the random baseline the learned rApp improves URLLC on-time delivery by up to $20\%$ and eMBB satisfaction by up to $12\%$. Since the low-level D-SAC scheduler is identical across all arms, these differences reflect the learned high-level policy.

\begin{figure}[t]
  \centering
  \begin{subfigure}[t]{0.48\linewidth}
    \centering
    \includegraphics[width=\linewidth]{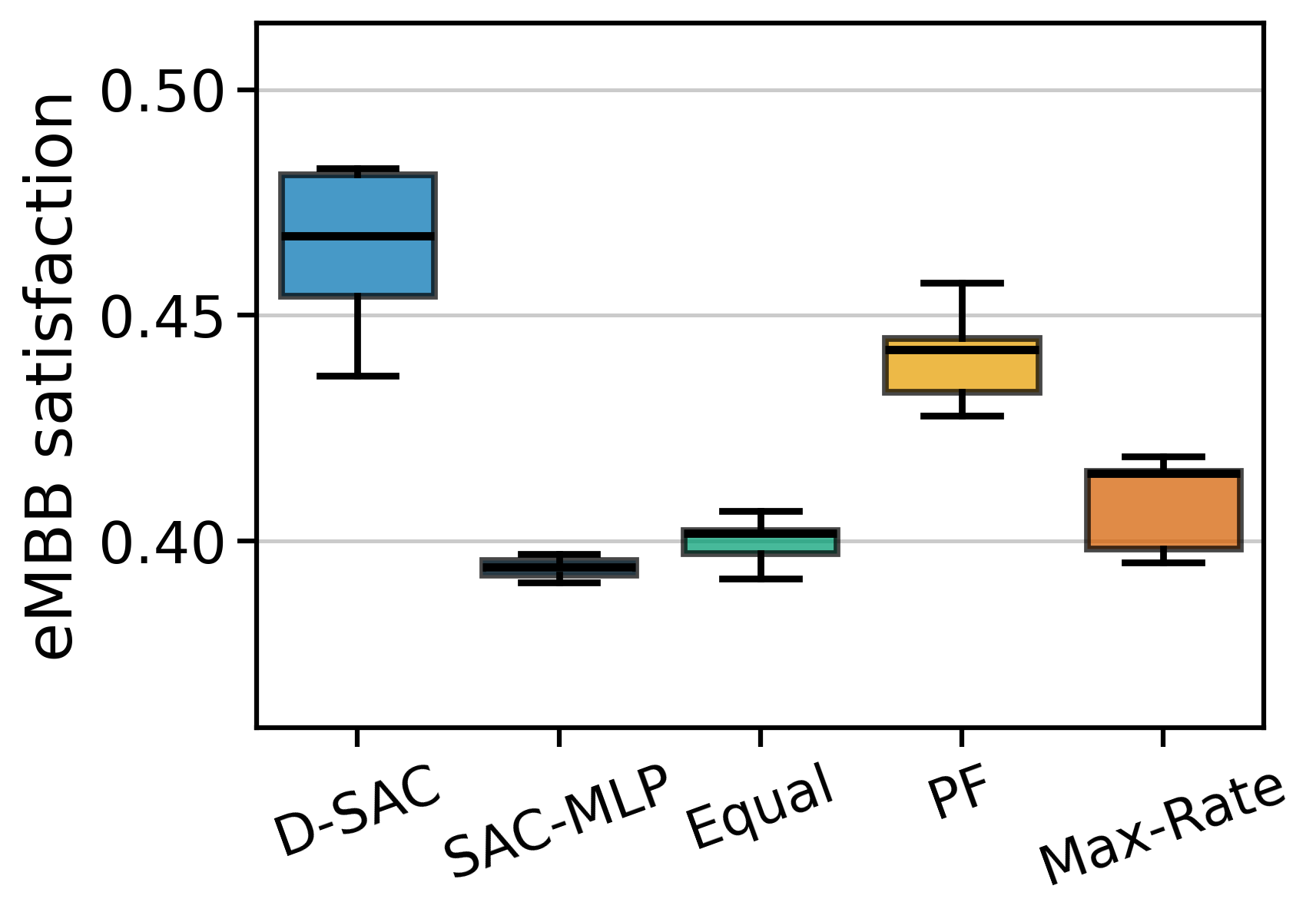}
    \caption{eMBB SLA satisfaction}
    \label{fig:results_xapp_embb}
  \end{subfigure}
  \hfill
  \begin{subfigure}[t]{0.48\linewidth}
    \centering
    \includegraphics[width=\linewidth]{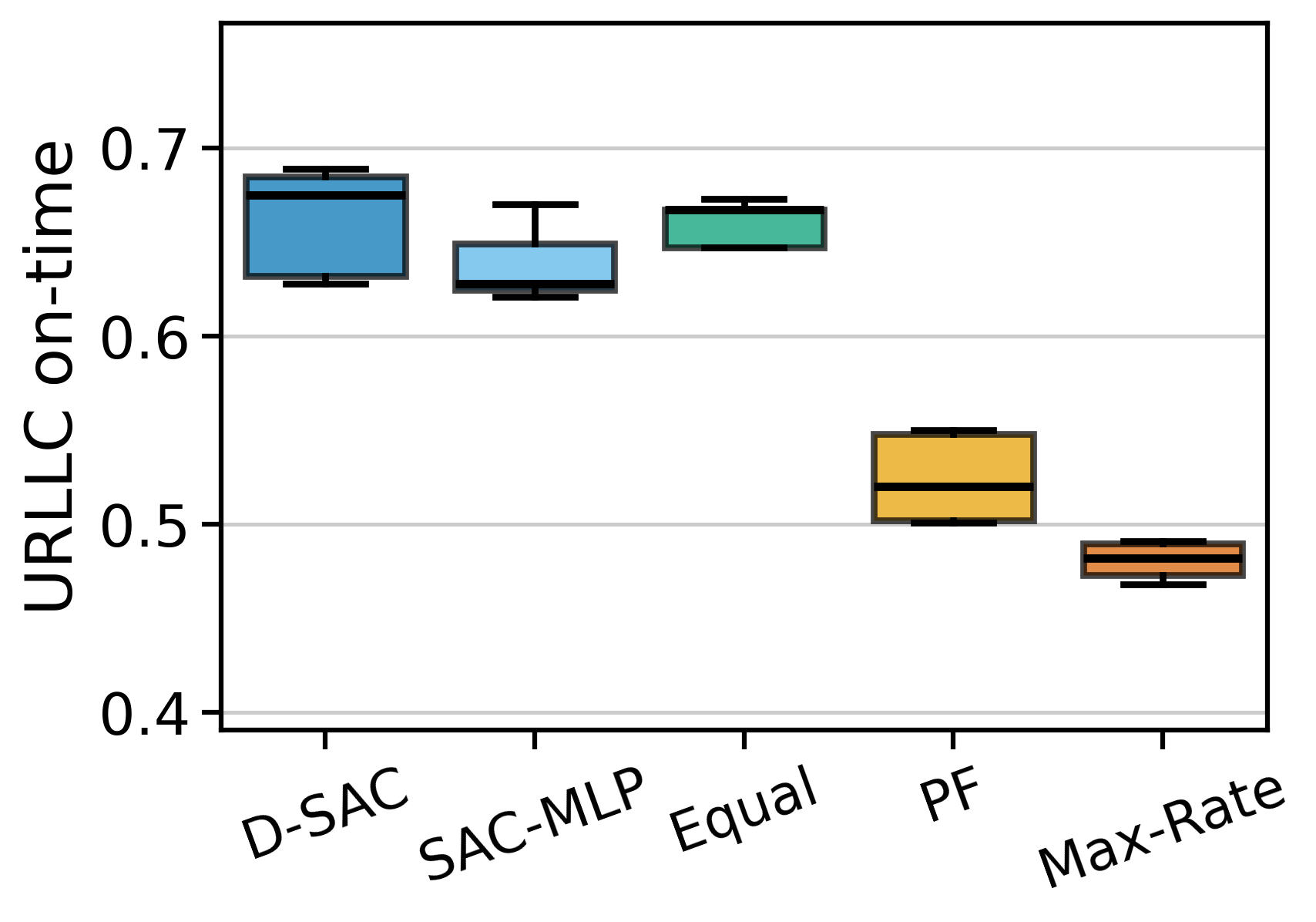}
    \caption{URLLC on-time delivery}
    \label{fig:results_xapp_urllc}
  \end{subfigure}
  \caption{Performance comparison; D-SAC outperforms all classical and learned baselines.}
  \label{fig:results_xapp}
\end{figure}

\subsubsection{Scheduler Comparison}
\label{subsec:results_xapp}
In Fig.~\ref{fig:results_xapp} we compare the proposed D-SAC scheduler with the classical proportional-fair (PF), Equal, and Max-Rate schedulers and a SAC-MLP, all running over the same learned rApp. All schedulers reach a comparable aggregate QoS utility ($0.695$--$0.724$), since the window-averaged eMBB rate exceeds $R_{\mathrm{target}}$, so the policies separate in the per-UE satisfaction metrics instead. The D-SAC achieves the highest eMBB SLA satisfaction ($0.463$, versus $0.429$ for PF, $0.403$ for Max-Rate, and $0.396$/$0.394$ for Equal/SAC-MLP) and the highest URLLC on-time delivery ($0.682$, versus $0.649$ for Equal and $0.524$/$0.481$ for PF/Max-Rate). The D-SAC leads the SAC-MLP on both metrics; the URLLC margin is narrow ($0.682$ versus $0.640$), but the decisive separation is on eMBB satisfaction, where the flat network collapses to near the Equal scheduler. This is the expected behavior of an index-tied parameterization: because per-UE traffic are randomized each episode, the UE index has no meaning, so without weight sharing the flat actor hedges toward a uniform split, which the Deepset model avoids.

\subsubsection{Service Tradeoff}
\label{subsec:results_tradeoff}
\begin{figure}[t]
  \centering
  \begin{subfigure}[t]{0.48\linewidth}
    \centering
    \includegraphics[width=\linewidth]{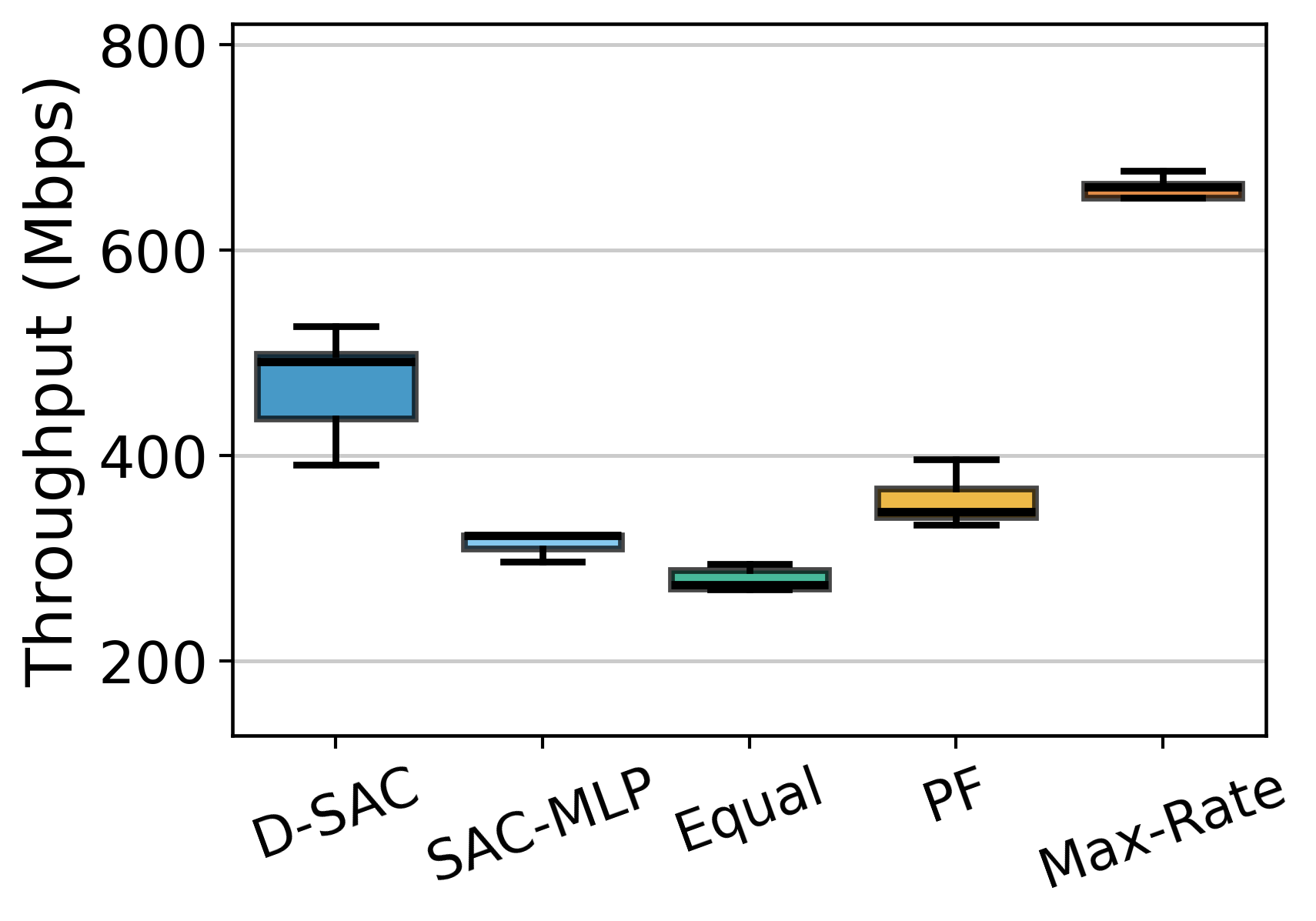}
    \caption{Throughput (Mbps)}
    \label{fig:results_tradeoff_throughput}
  \end{subfigure}
  \hfill
  \begin{subfigure}[t]{0.48\linewidth}
    \centering
    \includegraphics[width=\linewidth]{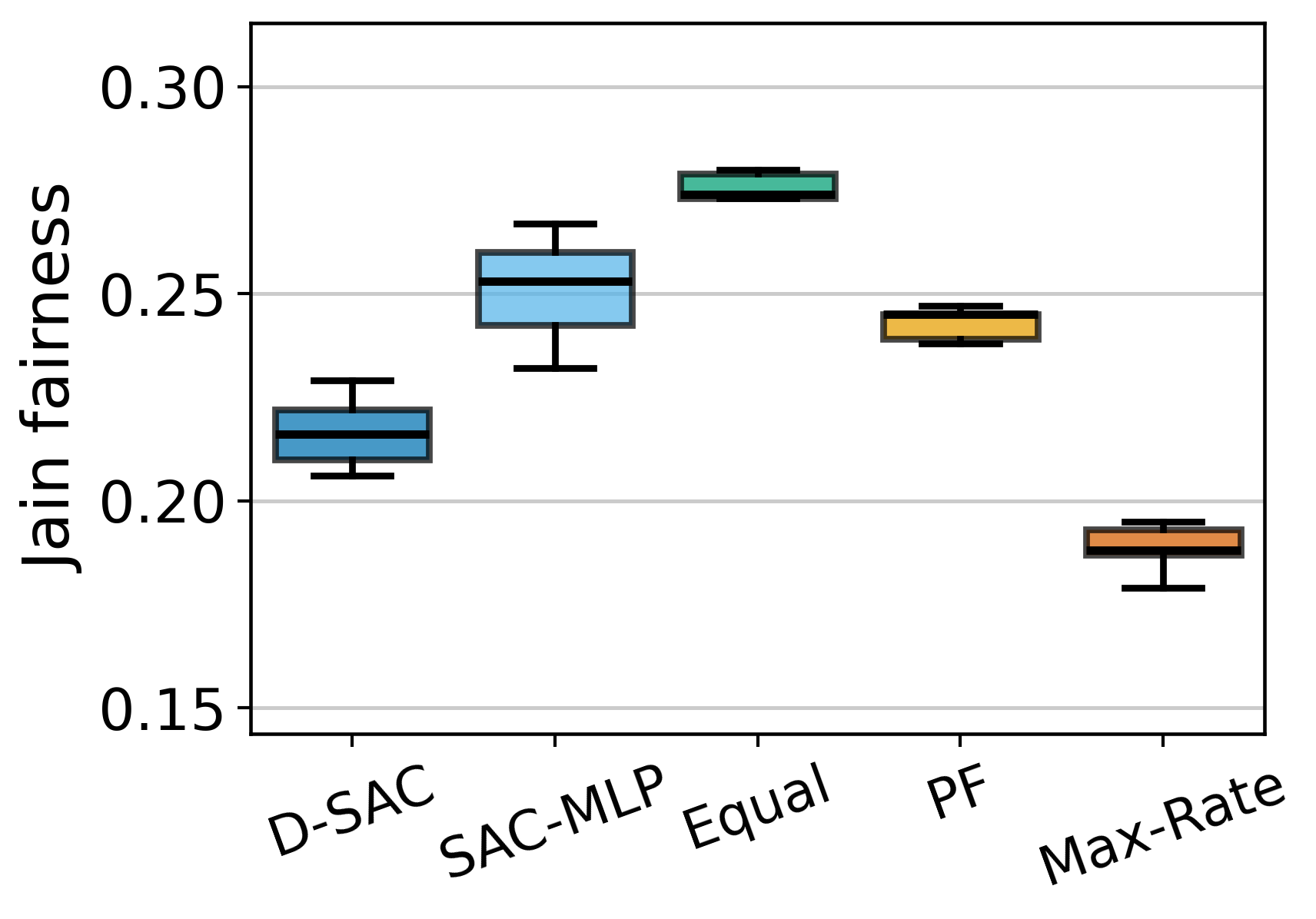}
    \caption{Jain's fairness}
    \label{fig:results_tradeoff_fairness}
  \end{subfigure}
  \caption{Throughput and Jain's fairness comparison across scheduling methods.}
  \label{fig:results_tradeoff}
\end{figure}

Fig.~\ref{fig:results_tradeoff} shows the throughput and fairness that each scheduler pays for the service quality. Max-Rate greedily serves users with the best instantaneous channel, reaching the highest aggregate throughput at the cost of the worst URLLC on-time delivery and the lowest fairness, while Equal is the most fair but throughput-limited and leaves much eMBB demand unmet. The D-SAC deliberately trades rate fairness for per-UE SLA satisfaction, which is the right trade when the per-UE minimum-rate contracts are heterogeneous.

\subsubsection{Scalability and Robustness to the Number of Users}
\label{subsec:results_scaling}
\begin{figure}[t]
  \centering
  \begin{subfigure}[t]{0.49\columnwidth}
    \centering
    \includegraphics[width=\linewidth]{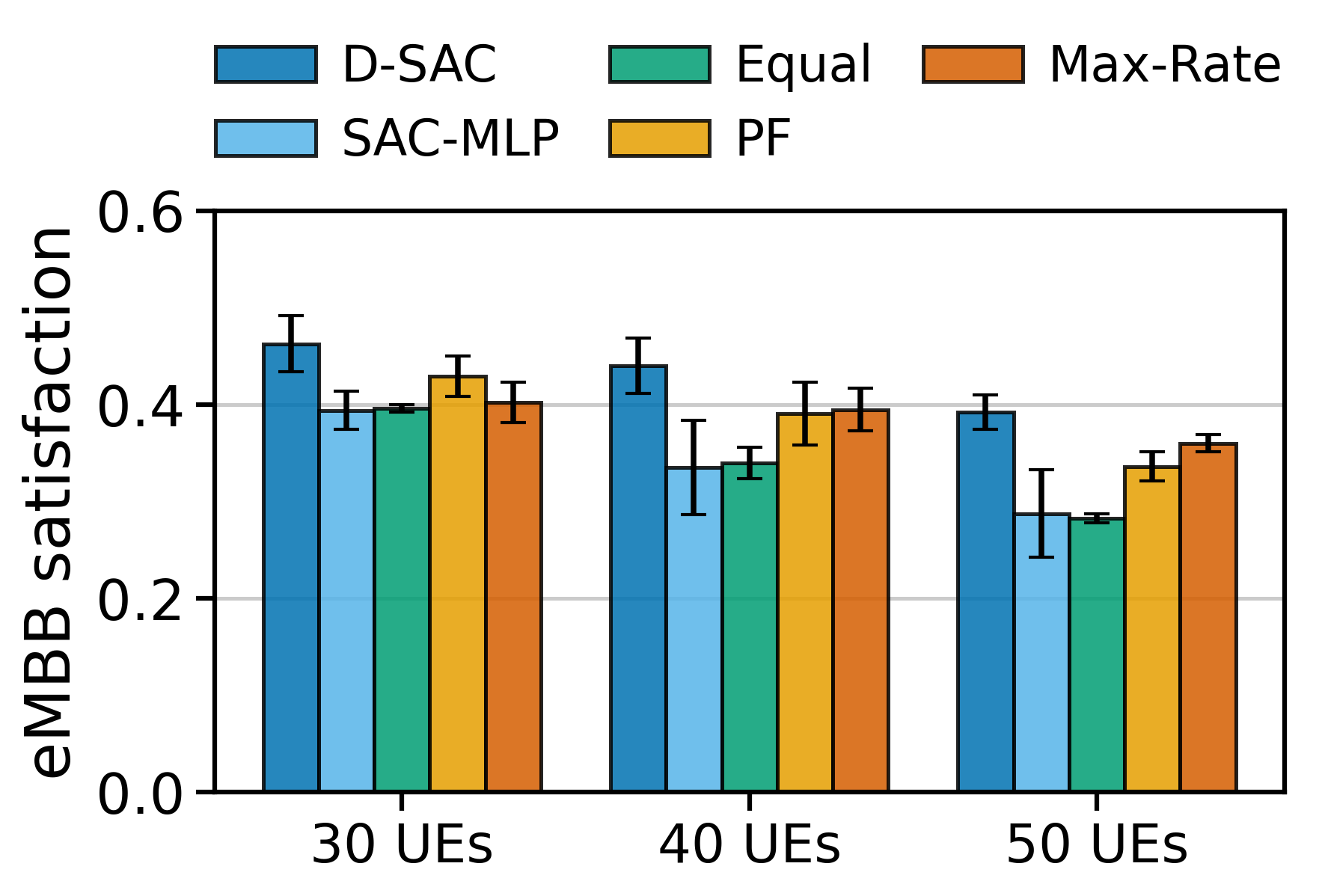}
    \caption{eMBB SLA satisfaction}
    \label{fig:results_scaling_embb}
  \end{subfigure}
  \hfill
  \begin{subfigure}[t]{0.49\columnwidth}
    \centering
    \includegraphics[width=\linewidth]{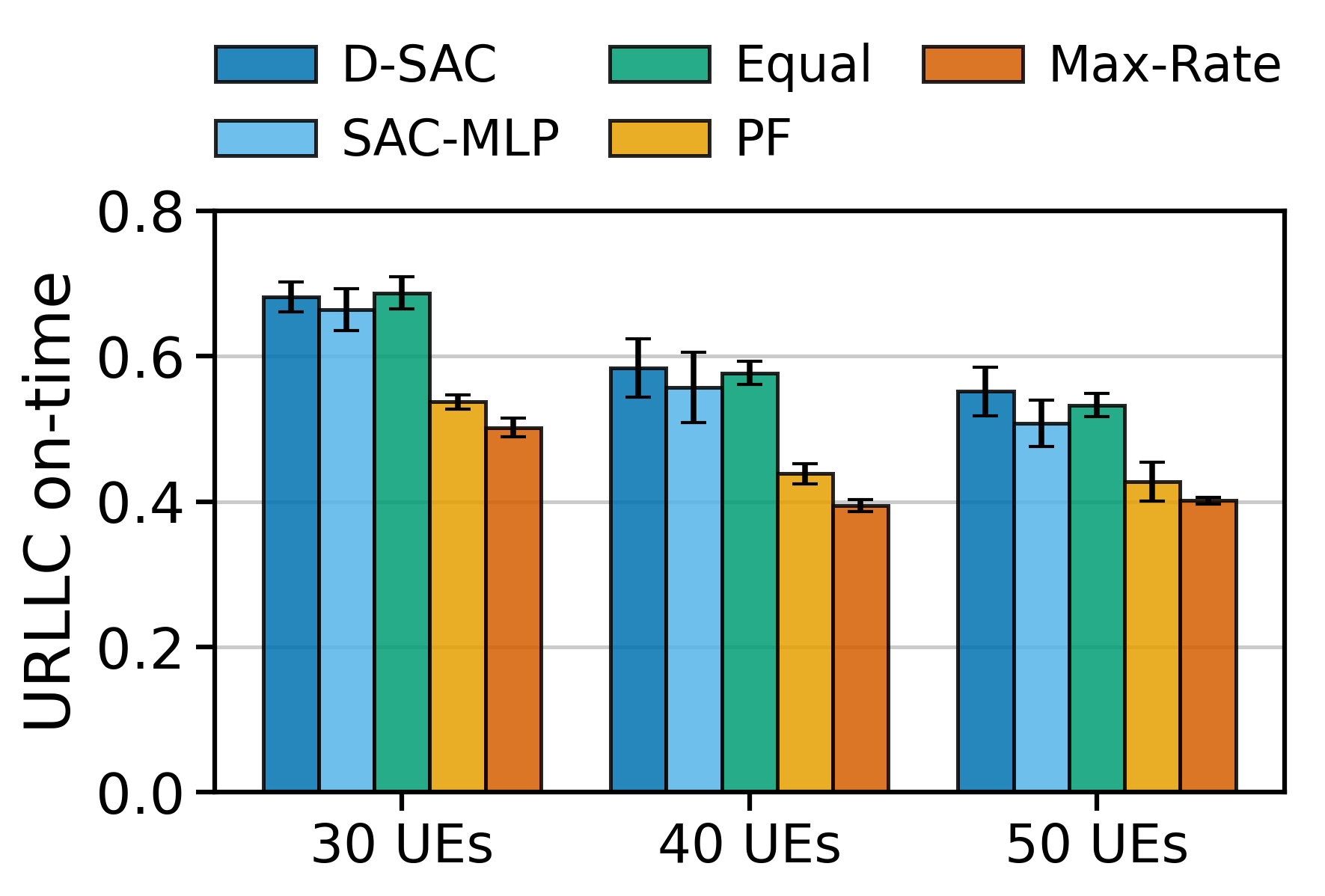}
    \caption{URLLC on-time delivery}
    \label{fig:results_scaling_urllc}
  \end{subfigure}
  \caption{Scalability performance. D-SAC leads on both.}
  \label{fig:results_scaling}
\end{figure}

Fig.~\ref{fig:results_scaling} shows that across $30/40/50$ UEs the D-SAC keeps the highest eMBB SLA satisfaction ($0.46$, $0.44$, $0.39$) and the highest URLLC on-time delivery ($0.68$, $0.58$, $0.55$) among all schedulers. The relevant result is that D-SAC retains the lead at every load without retraining: the actor is permutation-equivariant and parameter-shared across UEs, so a single trained model transfers to $40$ and $50$ UEs, whereas the flat SAC-MLP is tied to a fixed UE dimension and must be retrained separately at each count.

\section{Conclusion}
\label{sec:conclusion}
We proposed a hierarchical intelligent O-RAN controller for a tethered mmWave UAV-gNB serving mixed eMBB and URLLC traffic, in which a non-RT rApp sets the UAV placement and the slice PRB budget as a slow policy and a near-RT xApp performs per-user PRB allocation conditioned on that policy. Realizing the xApp as a permutation-equivariant DeepSets SAC scheduler gives it the natural symmetry of scheduling, unlike a flat MLP, and without the graph construction that GNN-based schedulers require. Evaluated on a ray-traced mmWave environment, the D-SAC improves eMBB SLA satisfaction by up to $17\%$ and URLLC on-time delivery by up to $42\%$ over all classical and learned schedulers, and is the only scheduler high on both service axes. The rApp improves URLLC on-time delivery by up to $20\%$ over random placement and surpasses a strong greedy-coverage heuristic, and the scheduler retains its lead from $30$ to $50$ users without retraining. Overall, the results show the benefit of jointly designing the two RIC control loops around per-UE  objectives.

\bibliography{ref}
\end{document}